\documentclass{PoS}
\usepackage{url}
\title{
\begin{picture}(0,0)(0,0)%
   \put(300,75){\makebox(0,0)[l]{\textnormal
{\normalsize OU-HET-881, KEK-CP-334 
}
}}%
\end{picture}%
Extracting the $\eta^\prime$ meson mass from gluonic correlators in lattice QCD}

\ShortTitle{Extracting the eta-prime meson mass from gluonic correlators}

\author{JLQCD Collaboration: 
        \speaker{H.~Fukaya}$^a$\thanks{E-mail: hfukaya@het.phys.sci.osaka-u.ac.jp},
        G.~Cossu$^{c}$,
        S.~Hashimoto$^{c,d}$,
        T.~Kaneko$^{c,d}$,
        \\
        \\
        \\
        \llap{$^a$}
        Department of Physics, Osaka University, 
        Toyonaka, Osaka 560-0043 Japan
        \\
        \llap{$^c$}
        High Energy Accelerator Research Organization (KEK),
        Tsukuba 305-0801, Japan 
        \\
        \llap{$^d$}
        School of High Energy Accelerator Science,
        The Graduate University for Advanced Studies
        (Sokendai),Tsukuba 305-0801, Japan
}

\abstract{Calculation of the $\eta^\prime$ meson mass is a notoriously difficult problem, as it requires evaluation of the disconnected diagram which is costly and noisy. In this work, we
use a gluonic operator to extract the eta-prime state after smearing the link
variables through the Wilson flow. With this choice, one can avoid a large
cancellation of pion contribution between the connected and disconnected diagrams. We obtain the $\eta^\prime$ meson mass on lattices with three different lattice spacings and two physical volumes, which allow us to estimate its continuum and large volume limits.}

\FullConference{The 33rd International Symposium on Lattice Field Theory\\
		14 -18 July 2015\\
		Kobe International Conference Center, Kobe, Japan*}

\begin{document}

\section{Introduction}

The $\eta^\prime$ meson is an interesting particle
among the low-energy hadrons in QCD.
While it would be a pseudo Nambu Goldstone (pNG) boson
if it is related to a spontaneous breaking of the axial $U(1)$ symmetry,
its mass $m_{\eta^\prime}$ ($=958$ MeV) is much heavier than
other pNG bosons like pion or kaon.
This $U(1)$ problem \cite{Weinberg:1975ui,'tHooft:1986nc} is one of the physical evidences
of the chiral anomaly, breaking of the symmetry at quantum level.

The axial $U(1)$ anomaly relates its violation to
the topological feature of the gluon fields.
More directly Witten \cite{Witten:1979vv} and Veneziano \cite{Veneziano:1980xs}
computed the mass of the
$\eta^\prime$ meson in the large $N_c$ 
(number of colors) limit, as a function of 
the topological susceptibility in QCD.

In QCD with $N_c=3$ and dynamical light quarks, however,
the argument of Witten and Veneziano is no longer valid.
It is not the $\eta^\prime$ meson but the pion 
that controls the topological susceptibility.
This was confirmed in our previous lattice QCD simulations where
we kept the chiral symmetry (nearly) exact
\cite{Aoki:2007pw, Hsieh:2009zz, Fukaya:2014zda}.
We found that the topological susceptibility 
is proportional to the light sea quark masses, 
consistent with the prediction from chiral perturbation theory,
$\chi_t = \frac{\Sigma}{\sum_i 1/m_i},$
where $\Sigma$ denotes the chiral condensate, and $m_i$ the $i$-th light quark mass. 
In particular, $\chi_t$ vanishes in the limit of massless up and down quarks,
reflecting the long-range dynamics of the pion field.

It is then interesting to ask what happens to the $\eta^\prime$ meson with $N_c=3$.
Since the effect of the anomaly is stronger than that of the large-$N_c$ limit,
the $\eta^\prime$ meson should be more sensitive to
the topological fluctuation of the gluon field, 
while it must be insensitive to $\chi_t$.
This implies a non-trivial double-scale structure in 
the topological excitation of gluon field:
it creates the $\eta^\prime$ meson at short distances,
while it is connected to the pion at long distances.

In this work, we perform $2+1$-flavor lattice QCD simulation, 
and show that the two-point function of the topological charge density at short distances
gives a mass consistent with the experimental value of the $\eta^\prime$ meson mass.
Since we have computed $\chi_t$ using the same correlation functions
(see \cite{Fukaya:2014zda} for the details),
our result clearly shows the double-scale structure
of the topological property of gauge fields.
Our results were already presented in a paper~\cite{Fukaya:2015ara}.
In this article, we review the main part of it.


Not only being theoretically interesting, but our work also 
provides a practically useful method to calculate the $\eta^\prime$ meson mass.
Direct lattice computation of the $\eta^\prime$ meson mass has been
challenging because of the disconnected diagram of quarks, 
which appears from the Wick contraction of the fermion
\cite{Kaneko:2009za,Christ:2010dd,Gregory:2011sg,Michael:2013gka,Ottnad:2015hva}.
This is numerically expensive and statistically noisy.

Using a gluonically defined operator of the flavor singlet pseudoscalar,
we can avoid the computational cost of
stochastically evaluating the disconnected diagram.
Our gluonic definition of the topological charge density does not
require any inversion of the Dirac operator.
Moreover, our method avoids the contamination from the pions.
In the conventional fermionic approach, one calculates both of 
the connected and disconnected diagrams of quark fields, 
both of which have the pion propagation and cancel with each other.
A large statistics is required for the cancellation 
before extracting the $\eta^\prime$ meson physics.
Since the purely gluonic definition of the topological charge density
\begin{equation}
  \label{eq:q}
  q(x) = \frac{1}{32\pi^2}\epsilon_{\mu\nu\rho\sigma}
  {\rm Tr}F_{\rm cl}^{\mu\nu}F_{\rm cl}^{\rho\sigma}(x),
\end{equation}
where $F_{\rm cl}^{\mu\nu}$ denotes the field strength tensor of
the gluon field defined through the so-called clover term made by 
four plaquettes, does not directly couple to the pions,
its correlator is free from the pion's fluctuation.

Note here that the sum of Eq.~(\ref{eq:q}) over the lattice volume
gives the global topological charge up to discretization effects.
In order to reduce the cut-off effects,
we {\it cool down} the link variables using the Yang-Mills (YM) gradient flow \cite{Luscher:2010iy}.
At a flow time $t$, it amounts to smoothing the gauge fields in a range
of the length $\sqrt{8t}$.
It was shown that the topological charge $Q$ defined through
(\ref{eq:q}) converges to an integer value at a sufficiently large
flow time \cite{Luscher:2011bx,Bonati:2014tqa}.
This smearing procedure eliminates short-distance noises and
also suppresses the noise at longer distances.\footnote{
  A similar method was tried in a quenched study
  to extract the ``pseudoscalar glueball mass''
  \cite{Chowdhury:2014mra}.
  Other types of smearings were tried in
  previous works to probe topological structure of the QCD vacuum 
  \cite{de Forcrand:1997sq,Horvath:2005cv,Ilgenfritz:2007xu, Alles:2007zz}.
}

In order to extract the $\eta^\prime$ meson mass, the YM gradient flow 
time should not be too long not to destroy the correlation of the $\eta^\prime$ propagation.
Assuming a Gaussian form of the smoothing effect,
Bruno {\it et al.} \cite{Bruno:2014ova} 
estimated the size of distortion of the correlator as
\begin{eqnarray}
\Delta \langle q(x)q(y)\rangle \sim e^{-(|x-y|/\sqrt{8t}-m_{\eta^\prime}\sqrt{8t})^2}\frac{m_{\eta^\prime} (8t)^{3/2}}{2\sqrt{\pi}|x-y|^2}.
\end{eqnarray}
In our analysis below, we use the reference flow 
time around $\sqrt{8t}=0.2$ fm for the fit range $|x-y|>0.6$ fm, 
for which we estimate the above correction to be less than 
$1$\% for $m_{\eta^\prime}\simeq 1$ GeV.

\section{Lattice setup}

In our simulations, the Symanzik gauge action and the $2+1$-flavor M\"obius domain-wall
fermion action are employed to generate gauge configurations \cite{Kaneko:2013jla, Noaki:2014ura, Cossu:2013ola}.
For the Dirac operator, three steps of stout smearing of the link variables are performed.
Our main lattice QCD simulations are performed on two different lattice volumes
$L^3\times T=32^3\times 64$ and $48^3\times 96$, for which
we set $\beta$ = 4.17 and 4.35, respectively.
The lattice cut-off $1/a$ is estimated to be 2.4~GeV (for $\beta=4.17$) and 3.6~GeV (for $\beta=4.35$),
using the input $\sqrt{t_0}=0.1465$ fm \cite{Borsanyi:2012zs}
where the reference YM gradient flow time $t_0$ defined 
by $t^2\langle E\rangle |_{t=t_0}=0.3$ \cite{Luscher:2010iy}
with the energy density $E$ of the gluon field, is used.
These two lattices have a similar physical volume size $L\sim 2.6$ fm.
We set the strange quark mass $m_s$ at around its physical point, 
and use 3--4 values of the up and down quark mass $m_{ud}$ for each $m_s$.
Our lightest pion mass is around 230 MeV with
our smallest value of $am_{ud}$ = 0.0035 at $\beta$ = 4.17.
In order to control the systematics due to finite volume sizes 
and lattice spacings, we also perform simulations on 
a larger lattice $48^3\times 96$ (at $\beta=4.17$ and $m_\pi \sim 230$ MeV),
and a finer lattice $64^3\times 128$ (at $\beta=4.47$ [$1/a\sim 4.5$ GeV] and $m_\pi \sim$ 285 MeV).
For each parameter set, we sample 500--1000 gauge configurations from 10000 molecular dynamics (MD) time.
We find that the residual mass in the M\"obius domain-wall fermion formalism is
kept smaller than $\sim$ 1~MeV \cite{Hashimoto:2014gta}
by choosing $L_s$ = 12 at $\beta$ = 4.17 and $L_s$ =  8  at $\beta$ = 4.35 (and 4.47).

On each generated configuration, we perform 500--1,000 steps of the
YM gradient flow (using the conventional Wilson gauge action) with a step-size $a^2\Delta t=$0.01. 
At every 20--30 steps, we store  $q(x)$ 
and measure its correlator using the Fast Fourier Transform (FFT) technique.

We find that the two-point function $\langle q(x)q(y)\rangle$ 
at our target distance $|x-y|\sim 0.7$ fm always
shows a shorter autocorrelation time than 10 MD time,
while that of the global topological charge, $Q=\sum_x q(x)$ 
is $O(100)$ or higher MD time at $\beta=4.35$.
This is a good evidence that the $\eta^\prime$ meson physics
is {\it decoupled} \cite{Schaefer:2010hu} 
from the physics of the global topological charge. 
In the following analysis, we estimate the statistical error
by the jackknife method after binning the data in 140--200 MD time.

\begin{figure*}[tbp]
  \centering
  \includegraphics[width=10cm]{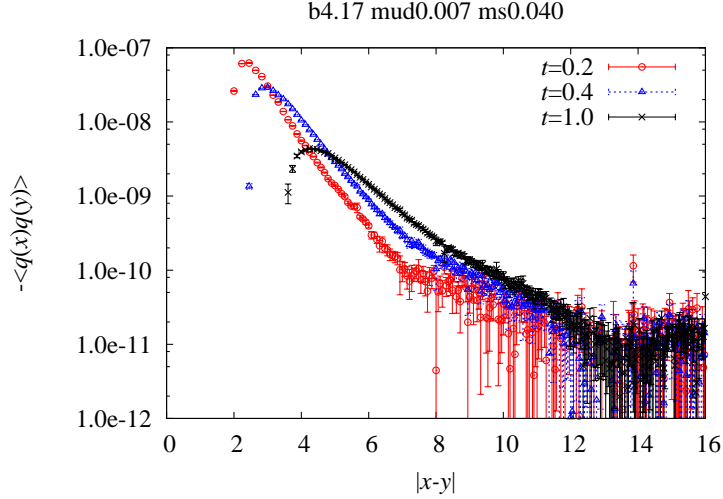}
  \caption{
    The correlator $-\langle q(x)q(y) \rangle$  at
    the flow times $a^2t$ = 0.2 (circles), 0.4 (triangles) and 
    1.0 (crosses). 
    Data at $\beta=4.17$, $am_{ud}=0.007$ and $am_s=0.040$ are presented.
  }
  \label{fig:corr}
\end{figure*}

\section{Numerical result}

Figure~\ref{fig:corr} shows the topological charge density correlator
$C(|x-y|)=-\langle q(x)q(y)\rangle$ at three different flow times. 
Using FFT, the rotationally symmetric data points are averaged.
As the flow-time increases, the statistical fluctuation of the
correlator becomes milder, while the region at small $|x-y|$ is distorted.
We therefore need to find a region of $t$ where the correlator
has sufficiently small noises 
while it is not spoiled by the smearing of the YM gradient flow.


To extract the mass $m_{\eta^\prime}$, 
we fit our data to the function of a single boson propagation:
\begin{eqnarray}
  f(r, m_{\eta^\prime})=A \frac{K_1(m_{\eta^\prime}r)}{r},
\end{eqnarray}
where $r=|x-y|$, 
$K_1$ is the modified Bessel function and $A$ is an unknown constant, which depends on the flow time $t$.
The fitting range, is determined by inspecting
a local ``effective mass'' $m_{\rm eff}(r)$,
a solution of 
$f(r+\Delta r, m_{\rm eff}(r))/f(r, m_{\rm eff}(r))
=C(r+\Delta r)/C(r)$, where we set $\Delta r=1/2$.
A reasonable plateau is found for $m_{\rm eff}(r)$ 
around $r\sim$ 8--12 ($> 0.6$ fm) at $t=1$ ($\sqrt{8t}\sim 0.2$ fm).

\begin{figure*}[tbp]
  \centering
  \includegraphics[width=10cm]{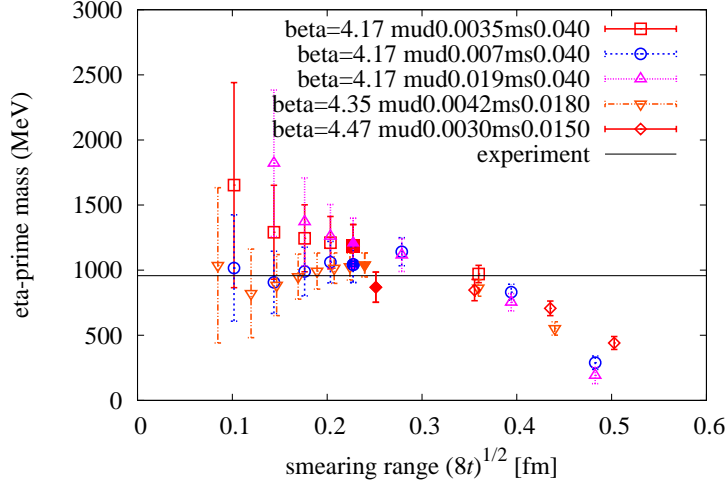}
  \caption{
    The flow-time dependence of the $\eta^\prime$ meson mass. 
    The data at various sea quark masses and $\beta$ values are shown,
    as specified in the legend. The filled symbols represent our data taken for the
    central values.
  }
  \label{fig:etapmass_vs_t}
\end{figure*}

Figure~\ref{fig:etapmass_vs_t} shows the obtained values of the $\eta^\prime$ meson mass 
as a function of $\sqrt{8t}$.
The data around $\sqrt{8t}\sim$ 0.2~fm are stable, while
a large distortion is found at larger smearing lengths $\sqrt{8t}\gtrsim$ 0.3~fm.
We take the data at $\sqrt{8t}=$ 0.2--0.25~fm 
(filled symbols in Fig.~\ref{fig:etapmass_vs_t}) for our results.

\begin{figure*}[tbp]
  \centering
  \includegraphics[width=10cm]{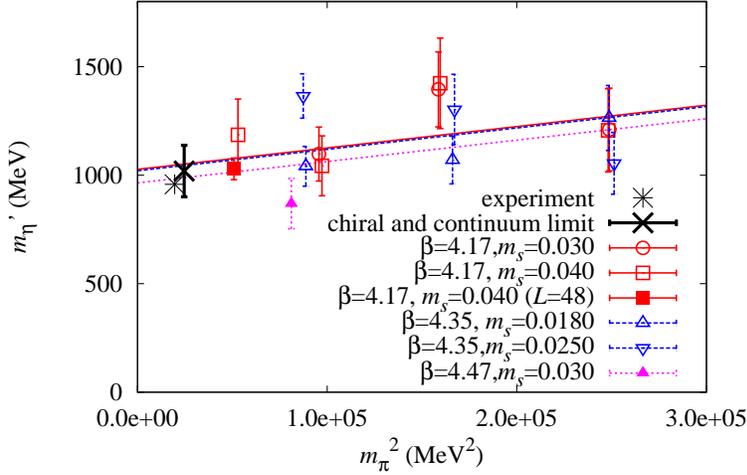}
  \caption{
    The extracted $\eta^\prime$ meson mass from each gauge ensemble.
    The three fit lines (representing the same linear fit function at three values of $a$)
    are shown for higher $m_s$'s at each $\beta$. 
}
\label{fig:etapmass}
\end{figure*}

We plot the results in Fig.~\ref{fig:etapmass} as a function of
the square of the pion mass $m_\pi$.
The results look insensitive to the quark masses, as well as to $V$ and $a$.
We therefore perform a global fit of our data 
to a linear function 
$m_{\eta^\prime}^{\rm phys} + C_a a^2 
+ C_{ud} [m_\pi^2-(m_\pi^{\rm phys})^2] + C_s[(2m_K^2-m_\pi^2)-\{2(m_K^{\rm phys})^2-(m_\pi^{\rm phys})^2\}]$,
where $m_{\eta^\prime}^{\rm phys},\; C_a,\; C_{ud}$, and $C_s$ 
are free parameters. Here, $m_{\pi/K}^{\rm phys }$ 
denotes the experimental value of the pion/kaon mass.
As shown by the lines (which are shown for higher $m_s$ only) in Fig.~\ref{fig:etapmass},
we find that our linear function fits the lattice data
reasonably well with $\chi^2$/(degrees of freedom) $\sim$ 1.6.

In Ref.~\cite{Fukaya:2015ara} we reported the study of
various systematic effects, including
the long auto-correlation of the global 
topological charge ($<1$\%),  
finite volume effects ($\sim 10^{-6}$),
the mixing with the $\eta$ meson ($\sim+5$\%),
and the chiral and continuum extrapolations ($\pm 8$\%).

Our final result at the physical point is
\begin{equation}
  m_{\eta^\prime} = 1019(119)(^{+97}_{-86})\;\;\;\mbox{MeV},
\end{equation}
which agrees well with the experimental value $m_{\eta^\prime}=957.78(6)$ MeV \cite{Agashe:2014kda}.
Here the first error is statistical and the second
is the systematic error from the mixing with the $\eta$ meson
and the chiral and continuum extrapolations (added in quadrature).

\vspace*{5mm}
We thank T.~Izubuchi, P.~de Forcrand, H.~Ohki and other 
members of JLQCD collaboration for fruitful discussions.
We also thank the Yukawa Institute for Theoretical Physics, 
Kyoto University. Discussions during the YITP workshop YITP-T-14-03 
on ``Hadrons and Hadron Interactions in QCD'' were useful to complete this work.
Numerical simulations are performed on IBM System Blue Gene Solution at KEK under 
a support of its Large Scale Simulation Program (No. 14/15-10). 
This work is supported in part by the Grand-in-Aid of the Japanese Ministry of Education 
(No.25287046, 25800147, 26247043, 26400259, 15K05065), and supported in part by MEXT SPIRE and JUCFuS.

\end{document}